\documentclass[conference]{IEEEtran}
\IEEEoverridecommandlockouts
\usepackage{cite}
\usepackage{amsmath,amssymb,amsfonts,mathtools}
\usepackage{algorithmic}
\usepackage{graphicx}
\usepackage{textcomp}
\usepackage{booktabs}
\usepackage{multirow}
\usepackage{listings}
\usepackage{adjustbox}
\usepackage{url}
\usepackage{makecell}
\usepackage{cite}
\usepackage{bigdelim}
\usepackage{breqn}
\usepackage[dvipsnames]{xcolor}
\usepackage{tikz}
\usepackage{pifont}
\usepackage{flushend}

\usepackage[T1]{fontenc}

\usepackage[labelformat=simple]{subcaption}

\usepackage[font=small,labelfont=bf]{caption}

\usepackage[dvipsnames]{xcolor}
\renewenvironment{thebibliography}[1]{
  \begin{oldthebibliography}{#1}
    \setlength{\itemsep}{1ex}
    \setlength{\parskip}{0ex}
}
{
  \end{oldthebibliography}
}

\usepackage{setspace}

\usepackage{titlesec}
\titlespacing*{\section}{0pt}{0.2\baselineskip}{0.1\baselineskip}
\titlespacing*{\subsection}{0pt}{0.2\baselineskip}{0.1\baselineskip}
\titlespacing*{\subsubsection}{0pt}{0.1\baselineskip}{0.0\baselineskip}

\allowdisplaybreaks

\def\BibTeX{{\rm B\kern-.05em{\sc i\kern-.025em b}\kern-.08em
    T\kern-.1667em\lower.7ex\hbox{E}\kern-.125emX}}

\begin{document}

\title{Faster Speech-LLaMA Inference with\\Multi-token Prediction}

\author{
    \IEEEauthorblockN{Desh Raj*, Gil Keren*, Junteng Jia, Jay Mahadeokar, Ozlem Kalinli}
    \IEEEauthorblockA{Meta, USA
    \\ \texttt{\{desh,gilkeren\}@meta.com}
    }
}

\maketitle

\begin{abstract}
% 100 to 150 words
Large language models (LLMs) have become proficient at solving a wide variety of tasks, including those involving multi-modal inputs. In particular, instantiating an LLM (such as LLaMA) with a speech encoder and training it on paired data imparts speech recognition (ASR) abilities to the decoder-only model, hence called Speech-LLaMA. Nevertheless, due to the sequential nature of auto-regressive inference and the relatively large decoder, Speech-LLaMA models require relatively high inference time. In this work, we propose to speed up Speech-LLaMA inference by predicting multiple tokens in the same decoding step. We explore several model architectures that enable this, and investigate their performance using threshold-based and verification-based inference strategies. We also propose a prefix-based beam search decoding method that allows efficient minimum word error rate (MWER) training for such models. We evaluate our models on a variety of public benchmarks, where they reduce the number of decoder calls by $\sim$3.2x while maintaining or improving WER performance.    
\end{abstract}

\begin{IEEEkeywords}
Large language models, Speech-LLaMA, speech recognition, inference.
\end{IEEEkeywords}

\section{Introduction}
\label{sec:intro}

% LLMs for speech recognition
Large language models have recently been shown to be able to perform a wide variety of tasks, including those requiring multi-modal understanding, through simple prompting-based approaches. For the case of speech inputs, it was found that attaching a small audio encoder and prompting an LLM (such as LLaMA) with these ``audial embeddings'' imparts general-purpose speech recognition (ASR) capabilities to the model~\cite{Fathullah2023PromptingLL,Fathullah2023AudioChatLlamaTG,Yu2023ConnectingSE,Tang2023SALMONNTG,Lai2023InstructionFollowingSR}. Since the LLM decoder in such models is trained on a web-scale text, the resulting models are usually found to be robust in transcribing multi-lingual speech.  

% Slow inference issue and possible solutions
Nevertheless, the use of LLM-based decoders for ASR requires the model and its transient states to be loaded into computing memory for each generated token in the auto-regressive decoding process. Unlike traditional ASR models like transducers~\cite{Graves2012SequenceTW} and attention-based encoder-decoders (AEDs)~\cite{Chan2015ListenAA}, these speech-LLMs comprise of a large and complex transformer-based decoder, making the generation process limited by memory bandwidth~\cite{Pope2022EfficientlyST} and preventing the model from making effective use of available compute. In effect, while the resulting ASR models usually outperform conventional systems, their perceived latency on user-facing applications are much higher.

% Methods to make LLM inference faster
Several methods have been proposed to speed-up the inference process for deep auto-regressive models (such as LLMs). These can be broadly categorized as \textit{model-based} and \textit{decoding-based} methods. Among model-based approaches, researchers have proposed probability density distillation~\cite{Oord2017ParallelWF}, reducing the KV-cache (for instance, using multi-query~\cite{Shazeer2019FastTD} and grouped-query~\cite{Ainslie2023GQATG} attention), subscaling~\cite{Kalchbrenner2018EfficientNA}, and quantization~\cite{Dettmers2022LLMint88M,Xiao2022SmoothQuantAA}. The dominant approach among the second category is speculative decoding~\cite{Leviathan2022FastIF,Chen2023AcceleratingLL}, which aims to execute several decoding steps in parallel, thus reducing the total number of steps required. Originally, this was realized by employing a smaller ``draft'' model, which speculates several subsequent tokens that the ``target'' LLM then verifies in parallel. This requirement was later alleviated by predicting multiple tokens from the target model itself, where each additional prediction head serves as the draft model~\cite{Gloeckle2024BetterF,Cai2024MedusaSL,Bhendawade2024SpeculativeSF}. This flavor of speculative decoding makes it analogous to block-wise parallel decoding for deep auto-regressive models~\cite{Stern2018BlockwisePD}.

% Contributions
In this paper, we incorporate the idea of multi-token prediction into decoder-only ASR models (such as Speech-LLaMA) to speed up inference. We investigate different model architectures and inference methods that enable multi-token prediction, and also propose new training objectives, such as a multi-head extension of sequence discriminative training. On experiments conducted using LibriSpeech and large-scale public multi-lingual benchmarks, we show that our proposed methods provide $\sim$3.2x reduction in the number of decoder calls, while maintaining or improving ASR performance.

\section{Preliminary: ASR with Speech-LLaMA}
\label{sec:speech_llama}

\begin{figure}
    \centering
    \includegraphics[width=0.8\linewidth,trim={0 0 1cm 0},clip]{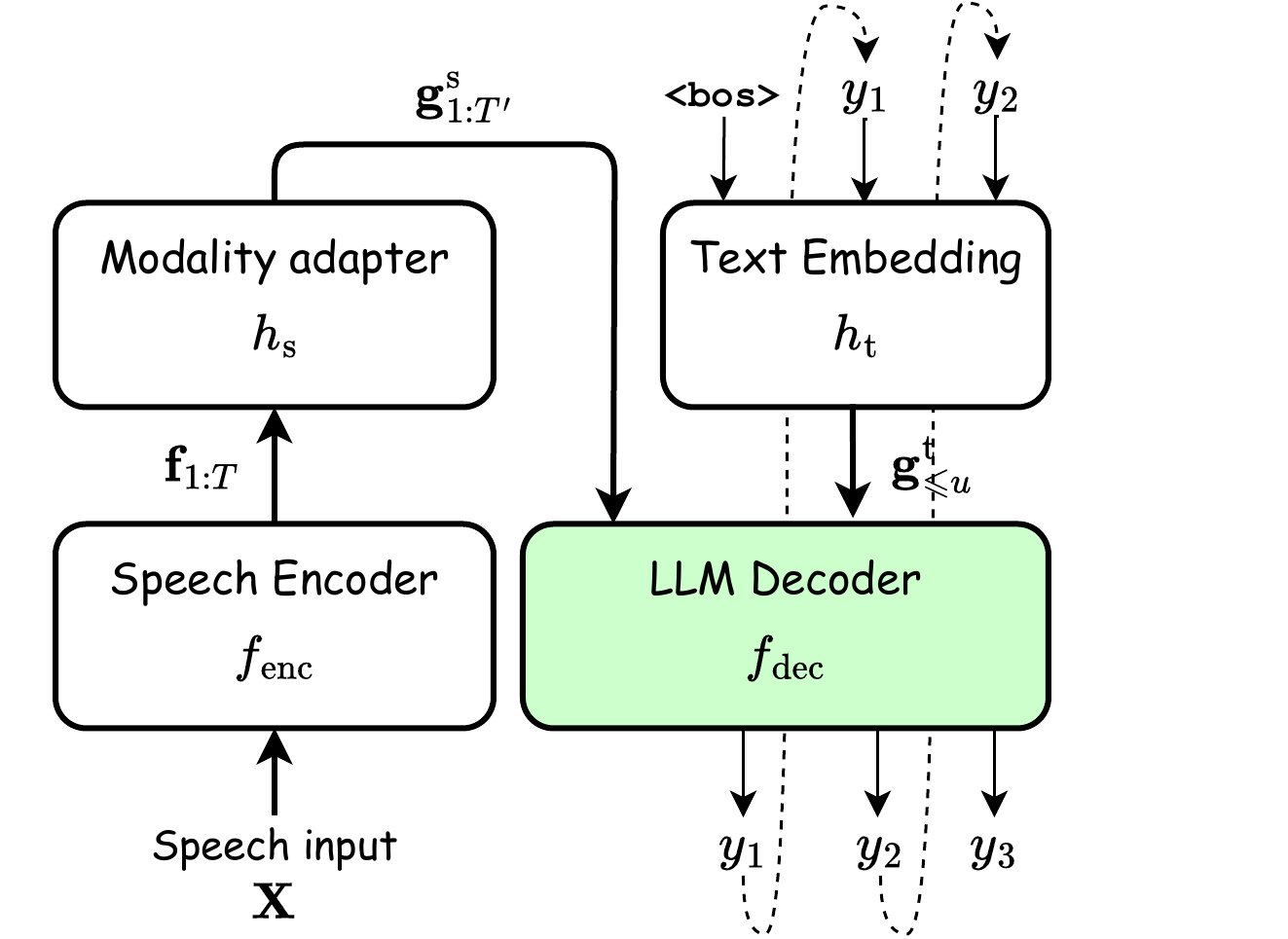}
    \caption{Overview of decoder-only ASR, e.g., Speech-LLaMA.}
    \label{fig:speech_llama}
\end{figure}

Given an utterance $\mathbf{X} = (\mathbf{x}_1,\ldots,\mathbf{x}_T)$, where $\mathbf{x}_t \in \mathbb{R}^d$ denotes audio features (usually log-Mel filterbanks), the task of an ASR model $f_{\mathrm{asr}}$ is to predict the most probable label sequence $\mathbf{y} = (y_1,\ldots,y_U)$, where $y_u \in \mathcal{V}$ denotes output units (e.g., word-pieces) including meta-tokens such as \texttt{<bos>} and \texttt{<eos>}. Mathematically,
\begin{align}
\mathbf{y} = f_{\mathrm{asr}}(\mathbf{X}) &= \mathrm{arg}\max_{\mathbf{y}} P(\mathbf{y}\mid \mathbf{X}) \label{eq:asr}\\
    &= \mathrm{arg}\max_{\mathbf{y}} \prod_{u=0}^{U-1} P(y_{u+1} \mid \mathbf{y}_{\leqslant u},\mathbf{X}). 
\end{align}

Decoder-only ASR models estimate $P(y_{u+1} \mid \mathbf{y}_{\leqslant u},\mathbf{X})$ using several parameterized components:
\begin{equation}
P(y_{u+1} \mid \mathbf{y}_{\leqslant u},\mathbf{X}) = f_{\mathrm{dec}}\left(\mathbf{g}^{\mathrm{s}}_{1:T'},\mathbf{g}^{\mathrm{t}}_{\leqslant u}\right), \label{eq:p_u}
\end{equation}
where $\mathbf{g}^{\mathrm{s}}_{1:T'} = h_{\mathrm{s}}(f_{\mathrm{enc}}(\mathbf{X}))$, and $\mathbf{g}^{\mathrm{t}}_{\leqslant u} = h_{\mathrm{t}}(\mathbf{y}_{\leqslant u})$.

Here, $f_{\mathrm{enc}}$ is a speech encoder that extracts high-dimensional representations $\mathbf{f}_{1:T}$ from the input speech $\mathbf{X}$, $h_{\mathrm{t}}$ converts discrete tokens $y_{\leqslant u}$ into embeddings $\mathbf{g}^{\mathrm{t}}_{\leqslant u}$, $h_{\mathrm{s}}$ is a modality adapter that sub-samples and projects $\mathbf{f}_{1:T}$ to $\mathbf{g}^{\mathrm{s}}_{1:T'}$ (in the same space as $\mathbf{g}^{\mathrm{t}}_{\leqslant u}$), and $f_{\mathrm{dec}}$ is an auto-regressive decoder. In practice, $f_{\mathrm{enc}}$ and $f_{\mathrm{dec}}$ are usually initialized from pre-trained models, such as a CTC-based ASR and an LLM, respectively~\cite{Fathullah2023PromptingLL}. The speech modality adapter $h_{\mathrm{s}}$ may be simple (e.g., frame subsampling with linear projection~\cite{Fathullah2023PromptingLL,Ling2023AdaptingLL}) or use non-linear contextualized modeling~\cite{Yu2023ConnectingSE,Chen2023SALMSL}. The decoder $f_{\mathrm{dec}}$ consists of $D$-dimensional self-attention blocks followed by a linear projection with softmax to obtain a distribution over $\mathcal{V}$, i.e.,
\begin{equation}
    f_{\mathrm{dec}}(\circ) \coloneq \mathrm{Softmax}\left(\mathbf{H}^{D\times V}\cdot f_{\mathrm{trf}}(\circ) \right), 
\end{equation}
where $V=|\mathcal{V}|$ and $f_{\mathrm{trf}}$ is the functional representation of the self-attention blocks. Fig.~\ref{fig:speech_llama} shows a diagrammatic representation of this model. For application in ASR, all model parameters are jointly trained on paired speech-text data $\{(\mathbf{X},\mathbf{y})\}^N$ using the cross-entropy loss as
\begin{equation}
    \mathcal{L}_{\mathrm{ce}} = -\frac{1}{N}\sum_{N}\sum_{u=1}^U \log P_{\theta}\left(y_{u+1}\mid \mathbf{y}_{\leqslant u},\mathbf{X}\right).
\label{eq:ce_loss}
\end{equation}
Additionally, a sequence discriminative training stage using the minimum word error rate (MWER) criterion (regularized with $\mathcal{L}_{\mathrm{ce}}$) is often employed. The MWER loss is given as
\begin{equation}
    \mathcal{L}_{\mathrm{mwer}} = \frac{1}{N}\sum_{N}\sum_{\mathbf{y}'\in \mathcal{B}(\mathbf{X})} \widehat{P}_{\theta}(\mathbf{y}'\mid \mathbf{X})\mathcal{W}(\mathbf{y},\mathbf{y}'),
\label{eq:mwer_loss}
\end{equation}
where $\mathcal{B}(\mathbf{X})$ represents the N-best list obtained using beam search and $\widehat{P}_{\theta}$ is the distribution re-normalized over $\mathcal{B}$.

The inference problem is to find $\mathbf{y}^{\ast} = \mathrm{arg}\max_{\mathbf{y}}P_{\theta}(\mathbf{y}\mid \mathbf{X})$. Since the output space is exponentially large, approximate search is usually through greedy or beam-search decoding, terminating when an \texttt{<eos>} token is generated.

\section{Multi-token Prediction}
\label{sec:method}

\begin{figure}
\centering
\begin{subfigure}{0.45\linewidth}
\centering
\includegraphics[width=\linewidth,trim={3cm 0 3cm 0},clip]{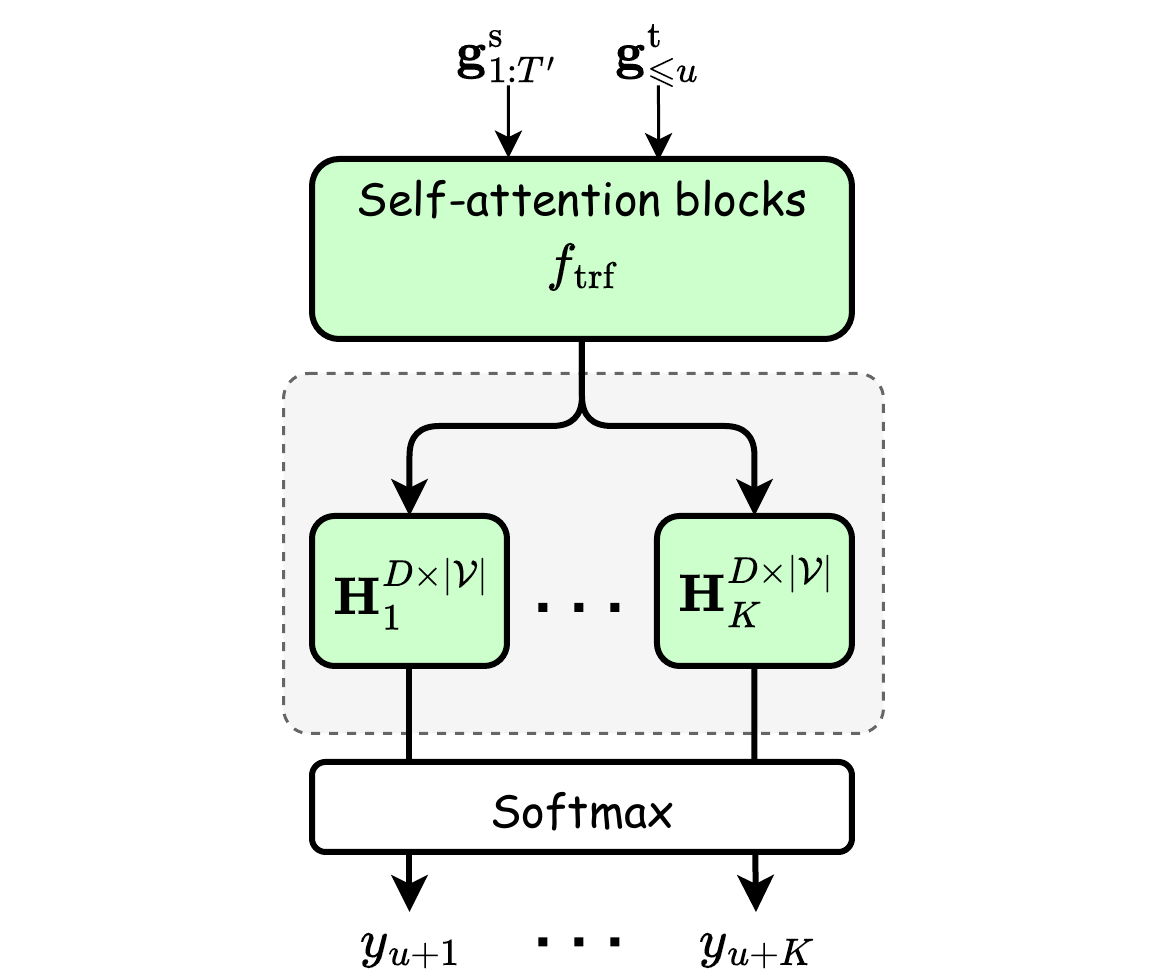}
\caption{}
\label{fig:medusa}
\end{subfigure}
\begin{subfigure}{0.45\linewidth}
\centering
\includegraphics[width=\linewidth,trim={0.4cm 0 0.4cm 0},clip]{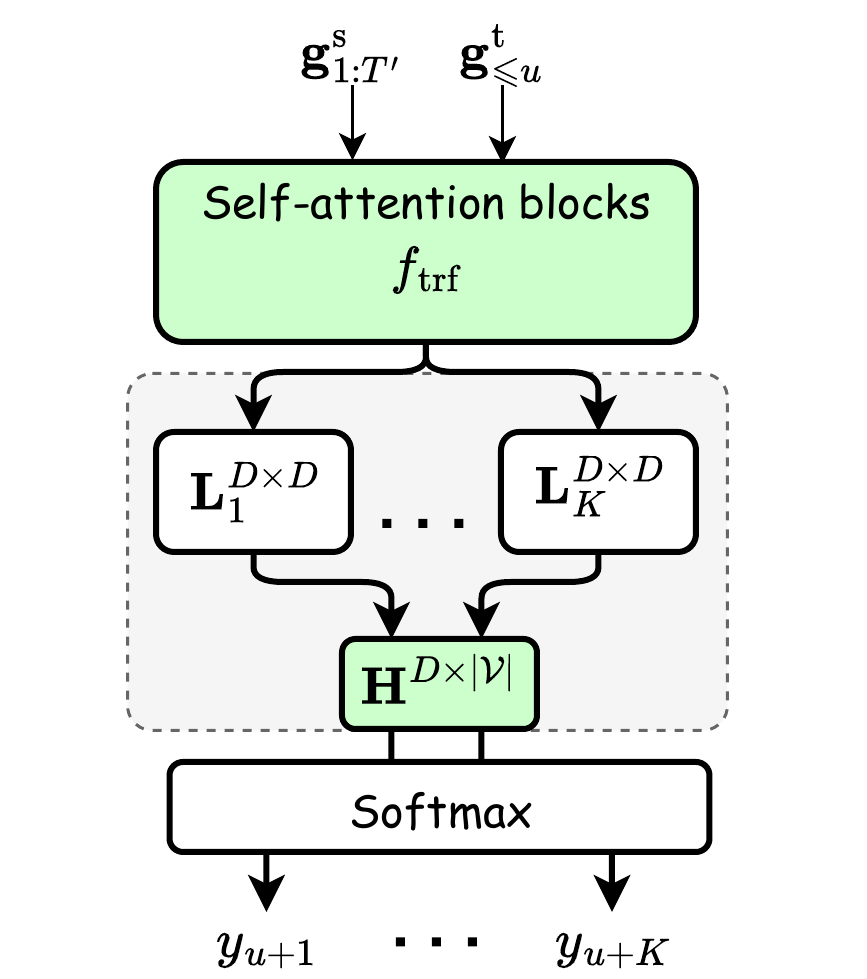}
\caption{}
\label{fig:latent}
\end{subfigure}
\caption{Architectures for multi-token prediction: (a) independent projection heads, and (b) latent-space expansion. The green-shaded blocks denote parameters initialized from the pre-trained LLM.}
\label{fig:arch}
\end{figure}

The model described in Section~\ref{sec:speech_llama} requires $U$ decoding steps to generate a sequence $\mathbf{y}$ of length $U$. This may be prohibitive when $U$ is large or when each decoder step is computationally expensive, as is the case when $f_{\mathrm{dec}}$ is an LLM. We conjecture that a complex decoder (such as an LLM) should be able to predict multiple tokens (say, $K$) in a single step of the decoding process, thus reducing the required number of decoding steps to $\frac{U}{K}$. Formally, we want a modified left-to-right factorization of \eqref{eq:asr} as
\begin{equation}
    P(\mathbf{y}\mid \mathbf{X}) = \prod_{u=0}^{\frac{U}{K}-1} P\left(\mathbf{y}_{uK+1:uK+K}\mid \mathbf{y}_{\leqslant uK},\mathbf{X}\right).
\end{equation}

We make a conditional independence assumption to approximate the $K$-token sequence probability as 
\begin{equation}
    P\left(\mathbf{y}_{u+1:u+K}\mid \mathbf{y}_{\leqslant u},\mathbf{X}\right) \approx \prod_{k=1}^K P_k\left(y_{u+k}\mid \mathbf{y}_{\leqslant u},\mathbf{X}\right),
\label{eq:medusa_factor}
\end{equation}
such that each $P_k\left(y_{u+k}\mid \mathbf{y}_{\leqslant u},\mathbf{X}\right)$ can be computed in parallel during the decoding step. We can use the same parametric estimate as in \eqref{eq:p_u} to compute the above probability by initializing and updating $K$ separate copies of $f_{\mathrm{dec}}$, one for each parallel token. However, this may result in a prohibitive increase in model size, especially when $f_{\mathrm{dec}}$ is large. To make the estimation tractable, we use a shared trunk $f_{\mathrm{trf}}$ for all the $K$ tokens, and only add $K$ independent parallel heads, each parameterized by a $D\times V$ matrix, i.e., $\mathbf{H}_1,\ldots, \mathbf{H}_K$, as shown in Fig.~\ref{fig:medusa}. This style of multi-token prediction was proposed in \cite{Gloeckle2024BetterF} and \cite{Cai2024MedusaSL} for faster LLM inference.

Despite sharing $f_{\mathrm{trf}}$, this strategy still adds $(K-1) DV$ extra parameters, which can be significant since modern LLMs are trained with large vocabularies, i.e., $D\ll V$. Inspired by \cite{Stern2018BlockwisePD}, we propose an alternate ``latent-space expansion'' formulation as shown in Fig.~\ref{fig:latent}. The idea is to factorize each projection head $\mathbf{H}_k$ into a full-rank $\mathbf{L}_k$ times an un-embedding matrix $\mathbf{H}$, where the latter is shared among all heads. This decomposition reduces the number of additional parameters to $KD^2$, making it independent of vocabulary size.

\subsection{Inference methods}
\label{sec:inference}

Due to the conditional independence assumption in \eqref{eq:medusa_factor}, the auxiliary heads are usually less accurate than the main head. As such, naively accepting all $K$ tokens for expanding the hypothesis at each step may result in performance degradation. Similar to previous work on multi-token prediction\cite{Stern2018BlockwisePD,Cai2024MedusaSL,Gloeckle2024BetterF}, we can use a ``predict-verify-accept'' paradigm to guarantee that the generated sequence is identical to the sequence that would be obtained under auto-regressive decoding.

\begin{enumerate}
    \item \textbf{Predict:} Get predictions for the $K$-token sequence, 
    $$\widehat{y}_{u+k} = \mathrm{arg}\max_{y_{u+k}}P_k(y_{u+k}\mid \mathbf{y}_{\leqslant u},\mathbf{X}), \forall~k =1,\ldots,K.$$
    \item \textbf{Verify:} Find the largest $\widehat{k}$ s.t.
    $$\widehat{y}_{u+r} = \mathrm{arg}\max_{y_{u+r}}P_1(y_{u+r}\mid \mathbf{y}_{< u+r},\mathbf{X}),~\forall r \leqslant \widehat{k}.$$
    \item \textbf{Accept:} Extend the hypothesis with verified tokens,
    $$\widehat{\mathbf{y}} \leftarrow \widehat{\mathbf{y}} \cup \{y_{u+1},\ldots,y_{u+\widehat{k}}\}.$$
\end{enumerate}

In practice, we combine the verification for the current step with the prediction for the next step, so that the decoder does not need to be invoked twice. We also experimented with weakening the verification criterion (in step 2) to top-$M$ selection, i.e.,
\begin{equation}
    \widehat{y}_{u+k} \in \text{top-}M_{y_{u+k}} P_1(y_{u+k}\mid \mathbf{y}_{<u+k},\mathbf{X}),
\end{equation}
where $M=1$ is equivalent to strict verification. While the verification-based inference allows our model to match the performance of auto-regressive decoding, it does not provide tight control over the performance-speed trade-off. As an alternate, we propose a novel \textbf{threshold-based selection} strategy by modifying the verification step as
\begin{equation}
   \widehat{k} = \max\{k: P_r(y_{u+r}\mid \mathbf{y}_{\leqslant u},\mathbf{X}) \geqslant \tau, ~\forall r \leqslant k\},
\end{equation}
where $\tau$ is a hyper-parameter.

\subsection{Training objective}
\label{sec:training}

For model training, we extend the cross-entropy loss in \eqref{eq:ce_loss} by summing over all $K$ predictions at each step, i.e.,
\begin{equation}
    \mathcal{L}'_{\mathrm{ce}} = -\frac{1}{N}\sum_{N}\sum_{u=1}^U \sum_{k=1}^K \alpha_k \log P_{\theta_k}\left(y_{u+k}\mid \mathbf{y}_{\leqslant u},\mathbf{X}\right),
\label{eq:medusa_ce_loss}
\end{equation}
where $\alpha_k$ is a hyper-parameter. We also perform sequence discriminative training using a modified version of the MWER loss in \eqref{eq:mwer_loss}. First, instead of regular auto-regressive decoding, we use thresholding-based multi-token prediction (\S~\ref{sec:inference}) to obtain the N-best hypotheses $\mathcal{B}(\mathbf{X})$ using beam search. To enable batched beam-search, we group the hypothesis by prefix length (instead of sequence length) during the forward calls, since this allows the transformer key-value states to be batched. Second, we compute the sequence log-probability by summing the token log-probabilities from the respective heads which generated them, i.e.,
\begin{equation}
    P_{\theta}(\mathbf{y}\mid \mathbf{X}) = \prod_{u=1}^U P_{\theta_{k(u)}}(y_u\mid \mathbf{y}_{\leqslant u-k(u)},\mathbf{X}),
\end{equation}
where $k(u)$ denotes the head that generated $y_u$. $P_{\theta}$ is then re-normalized over $\mathcal{B}$, similar to \eqref{eq:mwer_loss}.

\section{Experimental Setup}
\label{sec:setup}

\subsection{Datasets}
\label{sec:data}

We used the LibriSpeech (LS) dataset~\cite{Panayotov2015LibrispeechAA} for most of our ablation experiments w.r.t. modeling and inference. LS comprises 960h of training data sourced from English audio-books; results are reported on the \texttt{test-clean} and \texttt{test-other} sets, with some ablations on \texttt{dev-other} when specified. We also trained large-scale multi-lingual models on a combination of labeled data in English~(\texttt{en}), French~(\texttt{fr}), Italian~(\texttt{it}), German~(\texttt{de}), and Spanish~(\texttt{es}), sourced from Common Voice (CV)~\cite{Ardila2019CommonVA}, VoxPopuli (VP)~\cite{Wang2021VoxPopuliAL}, and Multi-lingual LibriSpeech (MLS)~\cite{Pratap2020MLSAL} datasets. We segmented and speed-perturbed the utterances from all sources, resulting in a total of 135k hours of paired data, and the durations of all subsets are shown in Fig.~\ref{fig:data}~(a). Since MLS (particularly the \texttt{en} subset) comprises a large fraction of the data, we performed over-sampling of the other sources during training, with effective ratios as shown in Fig.~\ref{fig:data}~(b). We also added $\sim$600k hours of in-house supervised data from these languages.

\begin{table}[t]
    \centering
    \begin{tabular}{cc}
    \adjustbox{max width=0.2\textwidth}{
    \begin{tabular}[b]{lrrr}
    \toprule
     & \textbf{\textcolor{Gray}{CV}} & \textbf{\textcolor{YellowOrange}{VP}} & \textbf{\textcolor{ForestGreen}{MLS}} \\
    \midrule
    \texttt{en} & 7.1 & 1.3 & 96.2 \\
    \texttt{fr} & 2.7 & 0.7 & 4.7 \\
    \texttt{it} & 0.9 & 0.3 & 1.0 \\
    \texttt{de} & 5.4 & 0.9 & 8.5 \\
    \texttt{es} & 1.8 & 0.5 & 2.8 \\
    \bottomrule
    \end{tabular}}
    &
    \includegraphics[width=0.5\linewidth,trim={0 2cm 0 1cm},clip]{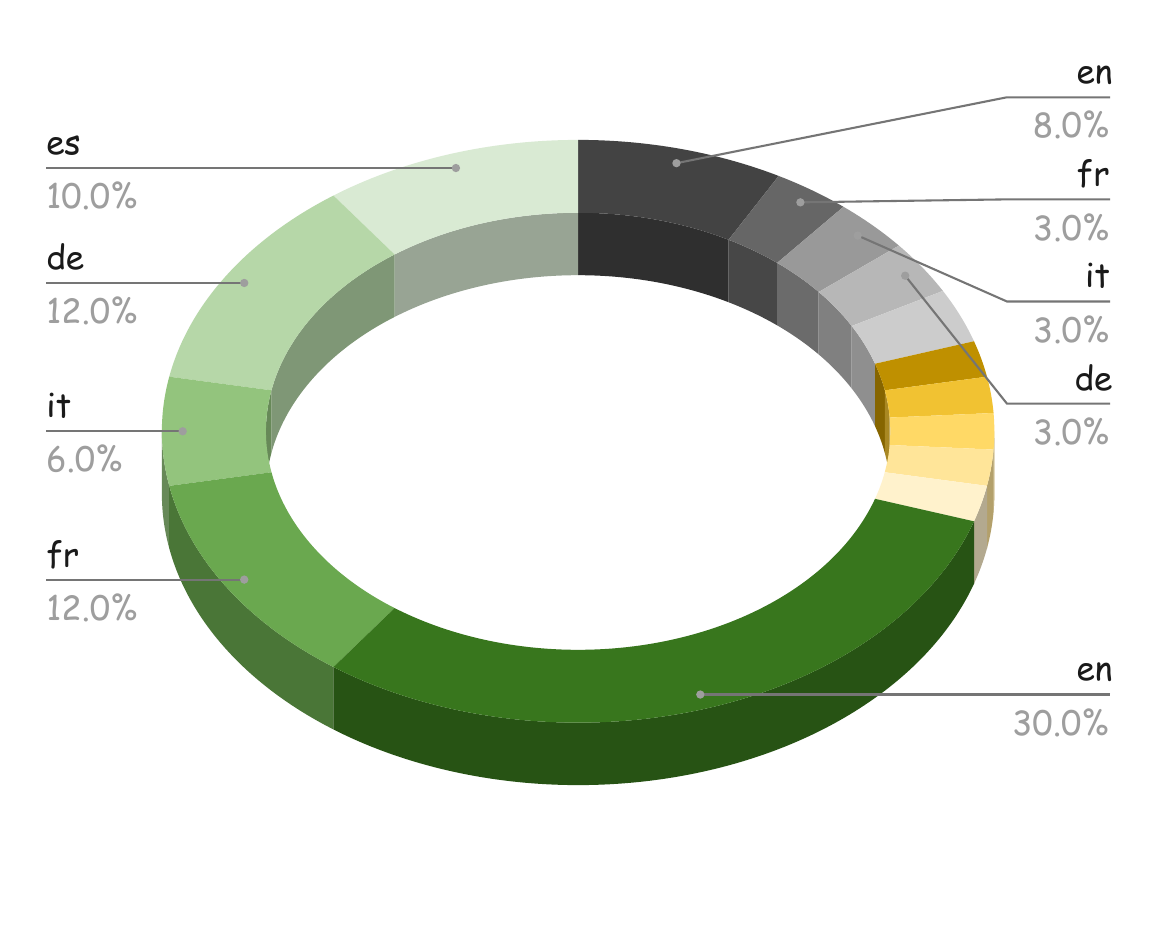} \\
    (a) Duration ($\times 10^3$ hours) & (b) Batch-wise sampling ratio
    \end{tabular}
    \captionof{figure}{Dataset statistics for large-scale multi-lingual training. The colors denote \textcolor{Gray}{Common Voice}, \textcolor{YellowOrange}{VoxPopuli}, and \textcolor{ForestGreen}{MLS}, respectively. Each VP subset is sampled in 2\% of the batches.}
    \label{fig:data}
\end{table}

\subsection{Evaluation metrics}
\label{sec:metrics}

We are interested in the performance of our systems on the ASR task, and the speed-up obtained during decoding. For the former, we report the well-known word error rate (WER) metric. For the latter, we compute the total number of forward calls made to the decoder as a ratio of the number of words in the reference and the hypothesis, and report their harmonic mean, i.e.,
\begin{equation}
    \eta = \frac{2\nu}{N_{\mathrm{ref}} + N_{\mathrm{hyp}}}, 
    \label{eq:eta}
\end{equation}
where $\nu$ denotes the number of decoder forward calls, $N_{\mathrm{ref}}$ and $N_{\mathrm{hyp}}$ are the number of words in the reference and hypothesis, and $\eta$ is the harmonic mean of the ratios (lower is better). We also report the decoder real-time factor (RTF) defined as the time spent in the decoder as a ratio of the audio duration. 

\subsection{Implementation details}
\label{sec:implementation}

We used a 125M distilled LLaMA with dimensionality $D=768$ and vocabulary of size 32k as the $f_{\mathrm{dec}}$ for all our experiments. A 100M, 512-dim VGG-Conformer was used as $f_{\mathrm{enc}}$ for LibriSpeech, and a larger 200M, 768-dim encoder for the multi-lingual ASR experiments. Speech inputs $\mathbf{X}$ are 80-dim log Mel filterbank features extracted using 25ms window with a stride of 10ms. The VGG convolutional layers perform 6x downsampling on the input, and a modality adapter $h_{\mathrm{s}}$ performs simple 3x stacking with linear projection. For cross-entropy training, we used a tri-stage scheduler with a peak learning rate (LR) of $10^{-3}$ warmed up for 20k steps. The models were trained for 200 epochs on LibriSpeech, and 10 epochs on the large-scale multi-lingual data. The hyper-parameters $\alpha_k$'s in \eqref{eq:medusa_ce_loss} were set to 1 for $k=1$ and 0.2 for $k>1$ based on preliminary experiments. For MWER training, we trained the models for 1 epoch with LR reduced to $5\times 10^{-5}$. We only used MWER training for the multi-lingual setup since it was found to degrade performance on LibriSpeech. For decoding, we used beam search with beam size 4 for all experiments. All implementation was done using an in-house extension of the PyTorch-based \textit{fairseq} toolkit.

\section{Results \& Discussion}
\label{sec:results}

\begin{table}[t]
    \caption{Comparison of multi-token prediction architectures and decoding methods in terms of ASR performance (\% WER) and the decoding speed, measured by $\eta$ \eqref{eq:eta} and the decoder RTF.}
    \label{tab:libri}
    \centering
    \adjustbox{max width=\linewidth}{
    \begin{tabular}{@{}lccccccc@{}}
    \toprule
    \multirow{2}{*}{\textbf{Model}} & \multirow{2}{*}{
    \begin{tabular}{@{}c@{}}
    \textbf{Decoding} \\ \textbf{method} 
    \end{tabular}
    } & \multicolumn{3}{c}{\texttt{test-clean}} & \multicolumn{3}{c}{\texttt{test-other}} \\
    \cmidrule(r{0.2em}){3-5} \cmidrule(l{0.2em}){6-8}
    & & \textbf{WER} & $\eta$ & \textbf{RTF} & \textbf{WER} & $\eta$ & \textbf{RTF} \\
    \midrule
    Speech-LLaMA & --- & 4.3 & 1.25 & 0.79 & 6.9 & 1.26 & 0.79 \\
    \midrule
    \multirow{3}{*}{
    \begin{tabular}{@{}l}
        Projection \\
        (4-head)
    \end{tabular}
    } & top-$1$ & 4.1 & 0.54 & 1.15 & 6.4 & 0.56 & 1.16 \\
    & top-$5$ & 4.0 & 0.43 & 0.81 & 6.4 & 0.45 & 0.85 \\
    & $\theta=0.8$ & 4.0 & 0.38 & 0.63 & 6.3 & 0.42 & 0.61 \\
    \midrule
    \multirow{3}{*}{
    \begin{tabular}{@{}l}
        Latent \\
        (4-head)
    \end{tabular}
    } & top-$1$ & 3.8 & 0.53 & 0.91 & 6.4 & 0.55 & 0.86 \\
    & top-$5$ & 3.7 & 0.42 & 0.63 & 6.3 & 0.44 & 0.63 \\
    & $\theta=0.8$ & \textbf{3.7} & \textbf{0.38} & \textbf{0.47} & \textbf{6.2} & \textbf{0.41} & \textbf{0.45} \\
    \bottomrule
    \end{tabular}}
\end{table}

\begin{figure}[t]
    \centering
    \includegraphics[width=0.65\linewidth]{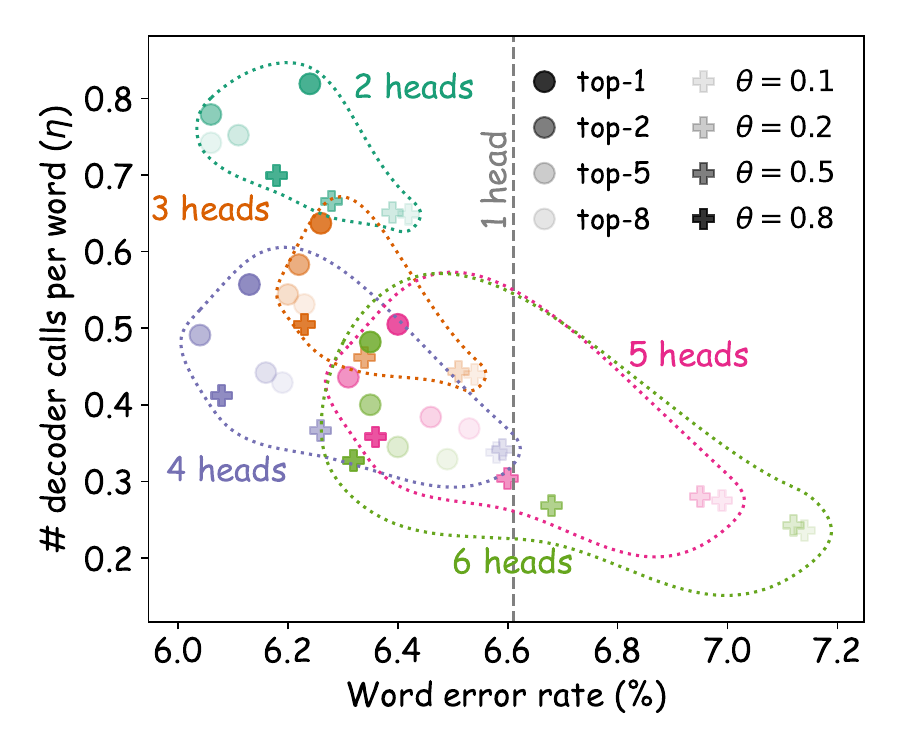}
    \caption{Plot of WER (\%) versus $\eta$ on LibriSpeech \texttt{dev-other} for different numbers of latent heads and inference methods. Each color represents a different model, and marker styles denote different decoding methods. Lighter markers denote weaker acceptance conditions, i.e., higher $M$ for top-$M$ and lower $\theta$ for threshold-based decoding. The single-head model has $\eta=1.26$.}
    \label{fig:wer_eta}
\end{figure}

\noindent
\textbf{Projection heads versus latent space expansion.} Table~\ref{tab:libri} shows a comparison between the two multi-token prediction architectures, ``projection'' and ``latent'' (\S~\ref{sec:method}), for $K=4$. In WER terms, both models outperformed the baseline Speech-LLaMA, with our best model providing 13.9\% and 10.1\% relative improvements on \texttt{test-clean} and \texttt{test-other}, respectively. For the same decoding method, both multi-token models made approximately the same number of decoder calls ($\eta$). However, the decoder RTFs obtained using the latent model (234M) were 22.7\% and 26.3\% smaller, on average, compared to the projection model (306M), on the two evaluation sets.

\noindent
\textbf{Effect of number of heads.} Next, we trained multiple Speech-LLaMA models with different numbers of latent heads, $K \in \{2,3,4,5,6\}$, and performed decoding on \texttt{dev-other} using the inference strategies described in \S~\ref{sec:inference}. We plot the WER versus $\eta$ for each such decoding run in Fig.~\ref{fig:wer_eta}. As $K$ increased, the average $\eta$ for the model reduced as expected, although it saturates after $K=5$. The WER performance was found to be consistent up to $K=4$, but worsened thereafter. Interestingly, almost all models improved WERs over the single-head baseline, which may be due to regularization effects from the auxiliary losses in \eqref{eq:medusa_ce_loss}. Overall, the 4-head model provided the best speed-performance trade-offs.

\noindent
\textbf{Comparison of inference strategies.} Among top-$M$ decoding methods (\tikz\draw[gray,fill=gray] (0,0) circle (.5ex);), $M=2$ was found to provide the best WERs for all models, indicating that auto-regressive decoding may be a suboptimal strategy for multi-head models. Threshold-based methods showed an almost linear response to changing $\theta$, with larger degradation in WER as $K$ increased. Unsurprisingly, top-$M$ methods resulted in better WERs (but worse $\eta$) than threshold-based approaches, but we did not see any common trend among all models with varying $M$.

\noindent
\textbf{Multi-lingual speech recognition.} Table~\ref{tab:multi_ling} shows the WER comparison between the baseline Speech-Llama and our 4-head latent model on the 3 benchmarks (CV, VP, and MLS). Unlike LibriSpeech, multi-token prediction degraded ASR performance with large-scale training by 4\% relative WER, on average. We believe that the 125M decoder may not have sufficient capacity to model the variability across domains and languages, and using a larger decoder should mitigate this degradation. MWER training improved WERs for the baseline and 4-head model by 3.6\% and 1.5\%, respectively. 

\begin{figure}[t]
    \centering
    \includegraphics[width=0.65\linewidth]{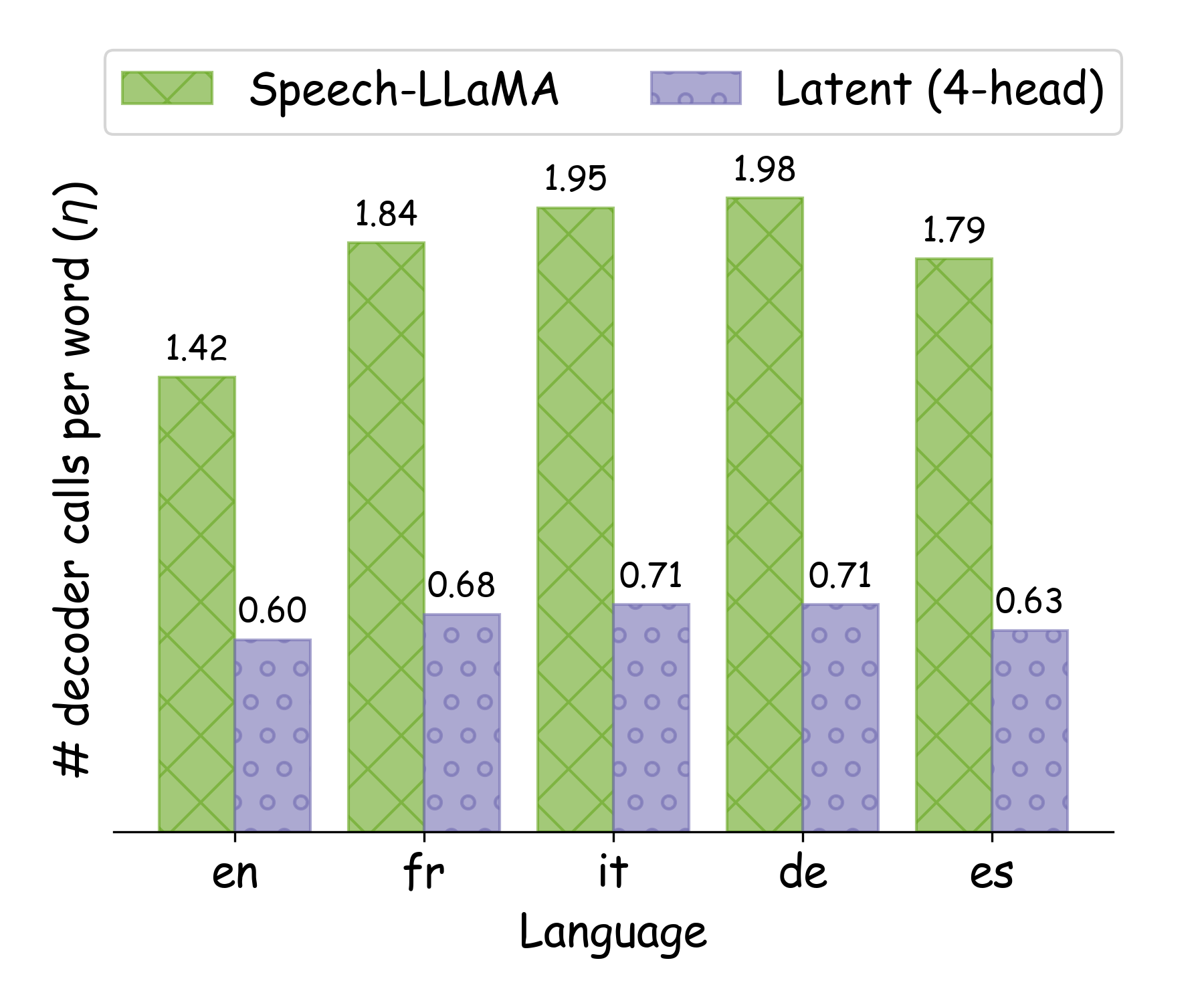}
    \caption{Comparison of number of decoder calls ($\eta$) for different languages, averaged over CV, VP, and MLS.}
    \label{fig:efigs_eta}
\end{figure}

\noindent
\textbf{Speed-up for different languages.} Despite the small increase in WER, we found that $\eta$ reduced significantly, by 62.8\% on average across all evaluation sets. In Fig.~\ref{fig:efigs_eta}, we compare the $\eta$ reduction for all five languages, for models trained without MWER loss. We found that multi-token prediction reduces the variability in $\eta$ across languages, which may be due to tokenization differences among languages.

\begin{table}[t]
    \centering
    \caption{Comparison of 4-head latent multi-token prediction model using threshold-based inference ($\theta=0.8$) vs. baseline Speech-LLaMA (in \% WER) on multi-lingual benchmarks.}
    \adjustbox{max width=\linewidth}{
    \begin{tabular}{@{}lrrrrrrrrrrrrrrr@{}}
    \toprule
    \multirow{2}{*}{\textbf{Model}} & \multicolumn{5}{c}{\textbf{\textcolor{Gray}{Common Voice}}} & \multicolumn{5}{c}{\textbf{\textcolor{YellowOrange}{VoxPopuli}}} & \multicolumn{5}{c}{\textbf{\textcolor{ForestGreen}{MLS}}} \\
    \cmidrule(r{0.2em}){2-6} \cmidrule(l{0.2em}r{0.2em}){7-11} \cmidrule(l{0.2em}){12-16}
    & \texttt{en} & \texttt{fr} & \texttt{it} & \texttt{de} & \texttt{es} & \texttt{en} & \texttt{fr} & \texttt{it} & \texttt{de} & \texttt{es} & \texttt{en} & \texttt{fr} & \texttt{it} & \texttt{de} & \texttt{es} \\
    \midrule
    Speech-LLaMA & 12.1 & 9.3 & 6.4 & 6.5 & 6.4 & 6.9 & 8.5 & 13.6 & 9.5 & 7.2 & 6.0 & 4.1 & 8.3 & 5.2 & 3.5 \\
    ~~+ MWER & 11.1 & 9.0 & 6.1 & 6.2 & 5.9 & 6.9 & 8.5 & 13.3 & 9.0 & 7.0 & 5.8 & 4.0 & 8.2 & 5.1 & 3.3 \\
    \midrule
    Latent (4-head) & 12.5 & 9.8 & 6.9 & 7.0 & 6.7 & 7.3 & 9.0 & 13.7 & 9.6 & 7.2 & 6.3 & 4.3 & 8.2 & 5.4 & 3.7 \\
    ~~+ MWER & 12.0 & 9.7 & 6.8 & 6.9 & 6.5 & 7.3 & 8.6 & 13.7 & 9.5 & 7.2 & 6.2 & 4.3 & 8.2 & 5.4 & 3.6 \\
    \bottomrule
    \end{tabular}}
    \label{tab:multi_ling}
\end{table}

\section{Conclusion}
\label{sec:conclusion}

We proposed to make Speech-LLaMA ASR inference faster by predicting multiple subsequent tokens at each decoding step. We compared two architectures to achieve this: using independent projections at the output, and using latent space expansion; we showed that the latter avoids significant increase in model size (resulting in lower overall RTF) while improving WER performance on LibriSpeech. We also investigated several inference strategies, demonstrating that a simple thresholding-based approach can offer better control over the performance-speed trade-off, compared to verification-based methods proposed earlier. Finally, our multi-lingual evaluations on CV, VP, and MLS indicated that our method reduces the variability in decoding speed among languages.

\noindent
\textbf{Acknowledgments.} We thank Jun Liu and Wei Zhou for helpful discussions, and Debjyoti Paul for preparing the multi-lingual datasets.

\bibliographystyle{IEEEbib}
\bibliography{refs}

\end{document}